# Superconducting radio-frequency cavity fault classification using machine learning at Jefferson Laboratory

Chris Tennant, Adam Carpenter, Tom Powers, Anna Shabalina Solopova, Lasitha Vidyaratne
*Jefferson Laboratory, Newport News, VA 23606*

Khan Iftekharuddin
*Old Dominion University, Norfolk, VA 23529*

We report on the development of machine learning models for classifying C100 superconducting radio-frequency (SRF) cavity faults in the Continuous Electron Beam Accelerator Facility (CEBAF) at Jefferson Lab. CEBAF is a continuous-wave recirculating linac utilizing 418 SRF cavities to accelerate electrons up to 12 GeV through 5-passes. Of these, 96 cavities (12 cryomodules) are designed with a digital low-level RF system configured such that a cavity fault triggers waveform recordings of 17 RF signals for each of the 8 cavities in the cryomodule. Subject matter experts (SME) are able to analyze the collected time-series data and identify which of the eight cavities faulted first and classify the type of fault. This information is used to find trends and strategically deploy mitigations to problematic cryomodules. However manually labeling the data is laborious and time-consuming. By leveraging machine learning, near real-time – rather than post-mortem – identification of the offending cavity and classification of the fault type has been implemented. We discuss performance of the ML models during a recent physics run. Results show the cavity identification and fault classification models have accuracies of 84.9% and 78.2%, respectively.

## I.   INTRODUCTION

The Continuous Electron Beam Accelerator Facility (CEBAF) at Jefferson Laboratory is a continuous-wave recirculating linac (see Fig. 1) [1, 2]. In September 2017 CEBAF completed an energy upgrade to extend its energy reach from 6 GeV to 12 GeV. To meet this milestone, 88 newly designed cavities (representing 11 cryomodules) were installed. Each cryomodule (comprised of eight 7-cell cavities) is capable of 100 MV energy gain and is regulated with an associated digital low-level RF system (LLRF). These are known as the C100 cryomodules.

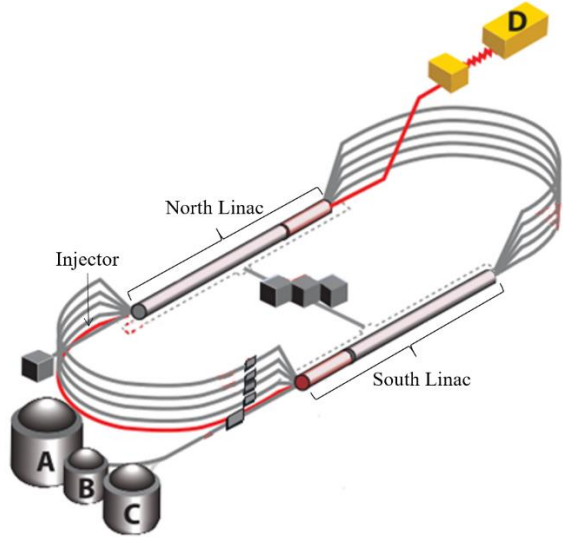

FIG. 1: Schematic of the CEBAF accelerator showing the four experimental halls (A, B, C, D) and indicating the locations of accelerating sections. The North and South Linacs are comprised of 25 cryomodules each. The five most downstream cryomodules of each linac are populated by C100s. The remaining two C100s are located in the Injector with one additional one in the North Linac.

As a nuclear physics user-facility, CEBAF aims to maximize beam availability to the four experimental stations. Currently, the largest contributor to short machine downtime trips (defined as events which are resolved in less than 5 minutes) are caused by superconducting radio-frequency (SRF) system faults. During 2019 there were an average of 4.1 RF trips per hour which translated to 58 minutes of beam time lost to RF recovery each day. The time lost in the experimental halls is effectively greater since data is discarded 30 seconds before the trip and 30 seconds after recovery. The C100 modules, in particular, were responsible for nearly a quarter of the downtime due to short trips across all accelerator subsystems. Presently the only operational parameter that operators can change in order to reduce the trip rate is the cavity gradient, which means that trip rate is inversely related to CEBAF's energy reach. Ideally, sufficient gradient overhead would exist to meet the demands of both tolerable trip rates and desired energy. In practice, there is little to no overhead and a compromise between trip rate and energy reach must be reached.

In order to better understand the nature and frequency of these faults we developed a new data acquisition system (DAQ) – by taking advantage of the digital LLRF system of the C100s – that simultaneously records waveforms of 17 different RF signals for each of 8 cavities in a cryomodule. This process is triggered when the LLRF system for any cavity in the cryomodule detects a fault condition. Typically, these waveforms are recorded at 5 kHz as compared to the conventional archived signals (up to 10 Hz). The stored waveform data provides sufficient resolution that, off-line, a subject matter expert (SME) can identify which cavity faulted first and the type of fault. However, in order to provide near real-time feedback to control room operators, an online system that utilizes machine learning to automatically identify (1) the cavity that faulted first and (2) the type of fault was developed and a prototype system implemented prior to the fall 2019 physics run. In this paper we outline the development of the various system components and report on initial results.

In Section II we further motivate this work and subsequent sections describe the various components that comprise the machine learning system. Specifically, Section III describes the data acquisition system, Section IV discusses the off-line analysis and the critical step of labeling the data, Section V details how the machine learning models were developed and implemented into a framework for online use. We conclude with results from the winter 2020 physics run and discuss future work.

## II.   MOTIVATION

The cavities in a C100 cryomodule have strong cavity-to-cavity mechanical coupling. When a cavity is turned off or the gradient is reduced, the change in length due to Lorentz force detuning leads to changes in the length of adjacent cavities. These effects are exacerbated by the mechanical resonances of the structure. The result is that when one cavity trips off, the remaining seven cavities are affected with several likely to trip from vibrational induced detuning. This leads to an operational problem with the C100 cavities, namely, answering the question "Which cavity tripped first?" Identifying the offending cavity with existing software and hardware is difficult to do automatically because secondary faults occur with delays on the order of milliseconds. While control room operators have access to the raw waveform data that is captured during a fault, correctly interpreting the signals requires a subject matter expert.

Developing machine learning models to automatically label fault events would immediately reduce the burden on SMEs and the time-consuming task of inspecting and labeling data manually. Having a labeled dataset is of interest at two different timescales. On the one hand, near real-time results after a fault event provide guidance to control room operators for appropriate responses (i.e. one type of fault might necessitate a reduction in cavity gradient, while another type of fault may indicate a larger problem and require intervention from system experts). On the other hand, using the aggregate statistics and breaking down faults according to cryomodule and cavity provides data driven guidance for maintenance and/or upgrade activities. We have already reaped the benefits of this latter kind of analysis. For example, post-mortem analysis of fault data from the fall 2018 physics run indicated three cryomodules in the South Linac (2L24, 2L25 and 2L26) were particularly susceptible to microphonic-based faults. This provided the impetus to perform microphonics hardening (installing tuner dampers) on these cryomodules during a down. As a result, in the following physics run, there was a reduction in microphonics-based trip rates, thereby allowing gradients to be increased in those cryomodules as well [3]. In another instance, results of manual fault classification uncovered a firmware bug which caused false trips of a type which represented approximately 40% of the faults. Additionally, results were used to monitor electronic quench fault types to better understand if there was a need to periodically thermally cycle C100 cryomodules to room temperature. Thus, both post-fault and post-run analysis by a SME is valuable. At present, only the latter is possible and is a laborious and time-consuming process.

## III.   DATA ACQUISITION

For each C100 cavity fault the newly developed data acquisition system synchronously acquires, timestamps and saves waveform records of 17 different RF signals from each of the eight cavities in the cryomodule. The data acquisition system is comprised of two primary components, the LLRF and EPICS control systems, along with a collection of high-level applications. These two components work together to generate and store the data for off-line analysis and inspection. An Altera field programmable gate array (FPGA) is the signal processing engine within the Jefferson Lab 12 GeV RF field control chassis (FCC) [4]. A waveform harvester was developed to capture RF time-series signals after a fault and write them to file for later analysis. Each of the 17 harvested waveform signals is 8192 points long. The trigger is set such that approximately 94% of the recorded data precedes the fault and 6% after. For standard sample rates of 0.20 ms per sample (5 kHz), this yields approximately 1535 ms of data before the fault. This pre-fault data provides valuable information about the root cause of the trip. After the data has been written to network storage, the harvester calls an external executable. Currently the executable runs the cavity and fault machine learning models and uploads the information to a centralized database via a waveform-specific web service. Finally, all waveform related data is kept online indefinitely, backed up to tape daily, and after 90 days is compressed to reduce online storage (see Fig. 2).

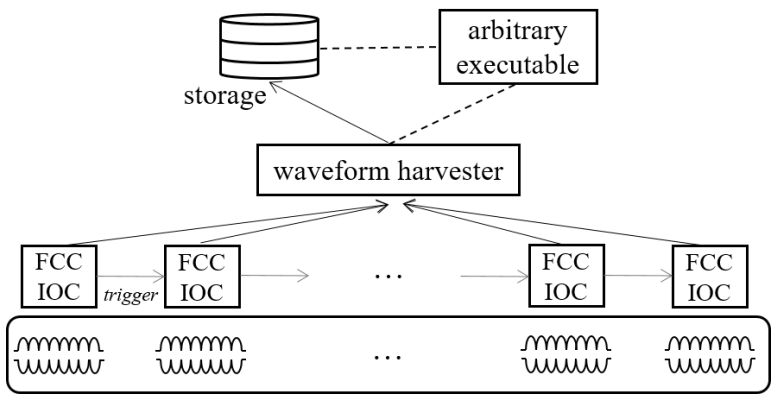

FIG. 2: A schematic showing the data generation and storage systems. A fault at a cavity triggers data collection at that cavity's FCC input/output controller (IOC), and the fault trigger is propagated to neighboring FCC IOCs associated with the cryomodule.

## IV. DATA ANALYSIS AND LABELING

Since deployment of the data acquisition system in early 2018, SMEs have labeled several thousand fault events from C100 cavities. That is, time-series signals from cavities have been assigned corresponding labels indicating the first cavity to trip off and the fault type. For the results in this paper we consider a dataset comprised of faults recorded during CEBAF beam operations between January 18, 2019 and March 9, 2020. The dataset contains a total of 2,375 labeled, C100 cavity fault events. Each event is comprised of 17 time-series signals per cavity for each of the eight cavities in the cryomodule. Each signal contains 1,638 ms of data sampled at 5 kHz. Details of each signal are listed in Appendix A. Figure 3 shows the distribution of faults by cavity, where the labels (1,2,3,4,5,6,7,8) correspond to unique cavity locations within a cryomodule and (0) corresponds to instances where all cavities fault, or were turned off, simultaneously. Figure 4 shows the distribution of faults by type.

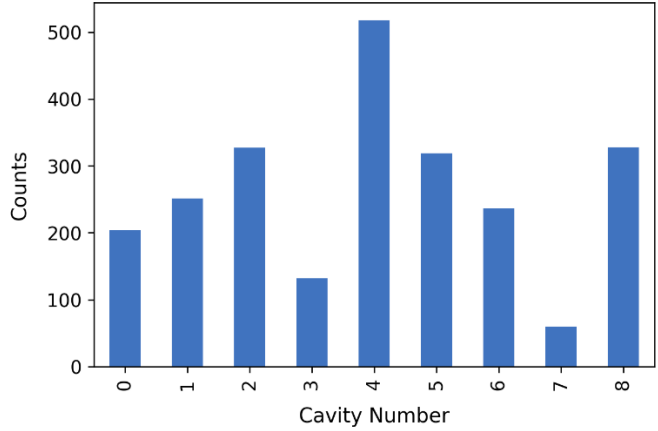

FIG. 3: Histogram showing the distribution of fault events by cavity identification. Numbers 1 through 8 correspond to unique cavity locations within a cryomodule and cavity 0 corresponds to instances where all cavities tripped off simultaneously.

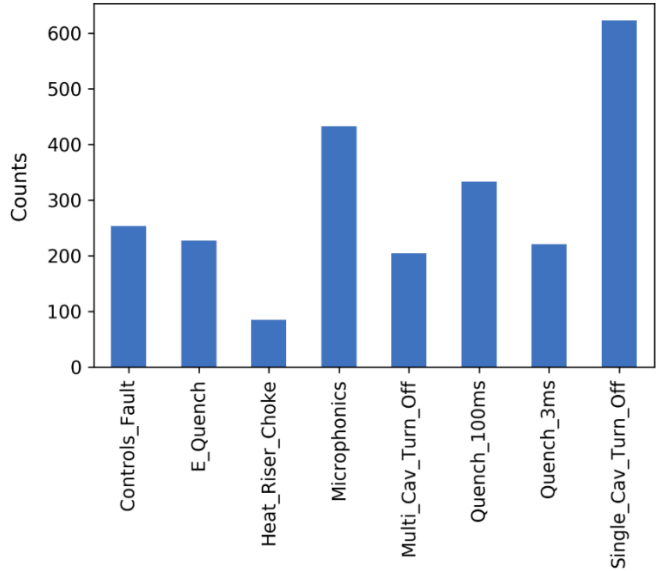

FIG. 4: Histogram showing the distribution of fault events by type. The data represents a total of 2,375 unique C100 cavity fault events.

Possible fault mechanisms [3] are; Microphonics, Quench_100ms, Controls_Fault, E_Quench, Quench_3ms, Single_Cav_Turn_Off, Heat_Riser_Choke, and Multi_Cav_Turn_Off. A microphonics fault is cause by detuning of the cavities, generally due to the vibrational modes of the cavity string. Because of the geometry of the structure the center of the string has a stiffer constraint and cavities 4 and 5 generally have higher frequency shifts and often will trip first. The 100 ms quench fault ("Quench_100ms") is assumed to be a thermal quench that is initiated in the end group which is conduction cooled. Detailed analysis done as part of the design process, indicated that the propagation time was 100 ms [5]. In these events the propagation of the quench is indicated by a single cavity having a substantial change in cavity frequency as indicated by a change in the relative phase between the forward RF power and the cavity field as measured with the field probe, also known as the detune phase. A 3 ms quench fault ("Quench_3ms") is similar in nature albeit with a faster shift in detune phase or an uncontrolled 1% reduction of cavity gradient with time scales on the order of 1 ms. An electronic quench fault ("E_Quench") refers to an event that involves a gas discharge in the cavity [6]. In these events the accepted theory is that electrons are stripped off of gas molecules and accelerated by cavity fields. In such events the stored energy in the cavity, as indicated by the measured gradient, is depleted in times on the order of 10 μs which is much faster than is possible with a thermal quench. More than 95% of the electronic quenches occur in cavities 1 or 8. This continues to be an area of investigation, though the working hypothesis is that it is due to gas that frozen out on the warm-to-cold beamline transition. An interlock fault ("Single_Cav_Turn_Off") occurs when a system interlock turns off a single cavity. In these faults there is no perturbation in the cavity RF signals prior to turning the RF off. A controls fault has two or three underlying mechanisms and can be attributed to the controls. One example is a single point perturbation in the system at the 5 kHz sample rates. Another common example is the several-kHz oscillation in the RF drive signal which occurs when the control loop phase offset is not set to the proper value. Heat riser choke faults ("Heat_Riser_Choke") occurs when the heat capacity of the vertical pipe between the helium vessel and the two-phase pipe is exceeded. The consequent pressure fluctuations affects the frequency of several cavities. Unlike a microphonics perturbation which builds up over several hundred milliseconds, a heat riser choke typically causes cavity frequency shifts that turn on in a step-wise manner. A multi-cavity turn-off fault ("Multi_Cav_Turn_Off") occurs when all 8 of the cavities are turned off simultaneously with no prior major variation in the RF signals. Typical drivers for such events are when the operators turn the zone off, when there is a beamline vacuum fault, or when there is a fault in the high-power amplifier, which is an 8-klystron assembly.

To better understand the labeling process, consider an example of a microphonics-induced fault. Microphonics describes time domain changes in the frequency of an SRF cavity, generally when the cavity has been perturbed externally. For cavities that have a large loaded-Q, changes in length of a few tens of nanometers cause sufficiently large frequency shifts that cause problems. Typically, cavities will vibrate at the modal resonance frequencies of the mechanical structure. For the C100 cryomodules the modes are the 9.5 to 10.5 Hz full string mode, 20 to 21 Hz half string mode, individual 40 to 45 Hz mode, and the 80 to 90 Hz tuner stack mode [7]. The cavity vibrations can also be excited by external sources such as 120 Hz from a vacuum pump which may be attached to the insulating vacuum vessel. Exciting the structure at any of these modal resonances amplifies the excitation and may cause the cavity to trip.

The RF signal that indicates detuning is the relative phase between the forward power and the cavity gradient signal. In CEBAF controls nomenclature this signal is labeled DETA2. Figure 5 depicts the waveforms for a microphonics-induced trip. As indicated by the DETA2 signal, the structure was vibrating in the 10 Hz full string mode. As the mode built up, the RF drive (GASK) and forward power signals (CRFP) increase to compensate for the perturbation. Eventually the control loop drive voltage signal for cavity 4 reached its maximum value of 10 and the drive phase got lost. In this instance the phase was such that it drove the cavity gradient down faster than the natural decay time of the system. Once cavity 4 tripped off, the remaining cavities were switched into self-excited loop (SEL) mode – a standard protocol with the C100 cavities used to speed up recovery from a trip.

This example illustrates both the domain knowledge required and the laborious nature of labeling fault events by hand. While this kind of post-mortem analysis is valuable, automating the process for real-time identification of the offending cavity and classification of the fault type would give control room operators actionable information to optimize C100 SRF performance.

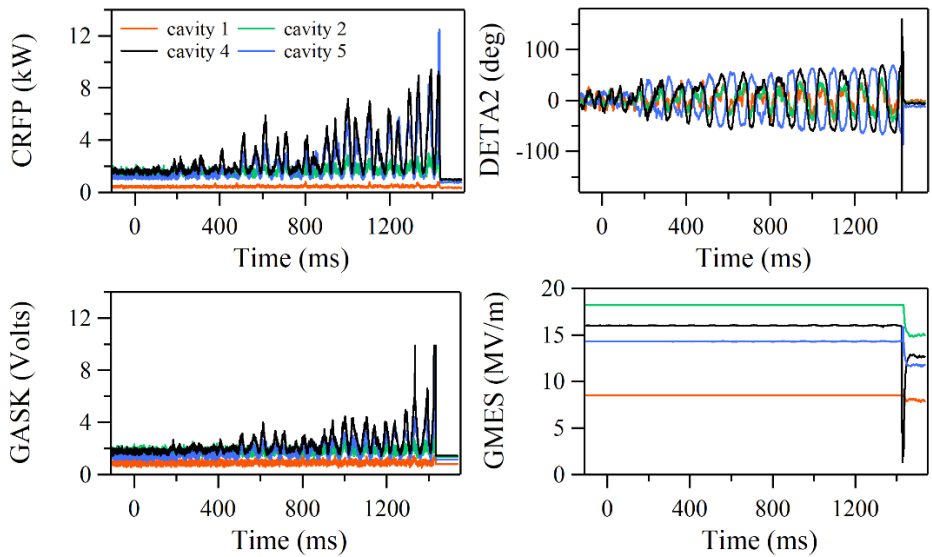

FIG. 5: Waveforms for a microphonics-induced fault. The plots display the forward power (top left), detune phase angle (top right), a digital signal proportional to the drive voltage (bottom left), and the measured gradient (bottom right). (Only 4 of the 8 cavities are plotted for clarity).

## V.  MACHINE LEARNING MODELS

Machine learning is particularly well suited for applications requiring pattern recognition [8]. The problem we are seeking to solve is one of sequence classification. More specifically, since we are dealing with a series of inputs over time, it is a problem of time-series classification. And because we have labeled examples to train the model, this represents a class of supervised learning. In this section we briefly outline how the models were developed.

### A.  Data preprocessing

The values in the raw signal waveforms exhibit large variations (orders of magnitude) between cavities and among signal types within a single cavity. To ensure that the objective function will be well behaved and to encourage faster convergence rates for gradient descent, we apply time-series standardization using the z-score technique on each waveform. The z-score function is shown in Eq. (1)

$$X_{norm} = \frac{X - \mu}{\sigma} \quad (1)$$

where $\mu$ and $\sigma$ are the mean and the standard deviation of $X$, respectively

## B. Feature extraction

The primary challenge for machine learning applications utilizing time-series data is feature extraction, in which statistical parameters (or features) are computed from the raw data signals. These features serve as an intermediate representation of the data and are used as model inputs. In order to reduce the computational load, features are computed for only 4 of the 17 recorded RF signals per cavity. These four signals are: (GMES, GASK, CRFP, DETA2) and were identified by SMEs as having the highest predictive power for both cavity identification and fault classification. The GMES signal is the calculated accelerating gradient in the cavity, GASK is a representation of a signal in the FPGA proportional to the RF drive voltage, CRFP is the measured forward power, and DETA2 is the relative phase between the RF voltage applied to the cavity and the electric field minus an offset phase (the offset phase is set such that DETA2 is zero when the resonant frequency of the cavity equals the machine reference frequency). Note that we utilize this set of common features for both cavity identification and fault classification models. In this configuration, we perform feature extraction once for each event, and maintain two independent classifiers trained for each task using the appropriate class labels. This strategy reduces the computational burden otherwise required to handle independent pipelines for each task separately.

These features can be a set of derived properties from the data that essentially convert the temporal classification to a more amenable static classification task [9]. These intermediate representations could be a set of simple statistics such as mean and variance, skewness, kurtosis, largest peak and number of zero crossings [10]. More descriptive features such as autoregressive coefficients [11], among others, may be required to obtain a more discriminatory representation of data. An autoregressive (AR) model is a linear predictive model where the next value of a time series is estimated based on previous values of the same time series using linear regression [12]. For instance, estimating $x^{(t)}$ based on $x^{(t-1)}$ is given as,

$$x^{(t)} = \varphi^0 + \varphi^1 x^{(t-1)} + \epsilon^t \qquad (2)$$

where $\varphi^0$ and $\varphi^1$ are regression parameters of the model, and $\epsilon^t$ is white noise. The model in Eq. (2) is regarded as a first-order autoregressive model (e.g. $AR(1)$) as the current estimate is based only on the previous value of the time series. The accuracy of the estimates may change based on the amount of historical data we use for the estimation with higher order AR modeling. A generalization of Eq. (2) for an AR model with order $k$ is given by

$$x^{(t)} = \varphi^0 + \sum_{i=1}^{k} \varphi^i x^{(t-i)} + \epsilon^t \qquad (3)$$

There are many ways to estimate the best parameters $\varphi$ for a given AR model, such as linear least squares methods. We utilize the ar_model.AutoReg [13] function in the Python-based "statsmodels" library for AR based feature extraction scheme. This function estimates the AR parameters $\varphi$ using the ordinary least squares method [12]. The process of fitting a AR($k$) model to a given time series results in a set of parameters, $\Phi = \{\varphi^0, \varphi^1, \dots, \varphi^k\}$, that best characterizes the waveform. These parameters become the features for our machine learning models. We select an AR model order of $k = 5$ to fit RF waveforms, as a compromise between feature length and model performance. This results in 192 features (8 cavities/event $\times$ 4 signals/cavity $\times$ 6 features/signal) for each fault event. The same features are used for both cavity identification and fault classification tasks.

### C. Model evaluation

Prior to training we split the data into a train (70%) and test (30%) set with stratification to ensure that the train and test sets have approximately the same percentage of samples of each target class as the complete set. We withhold the test data as the unseen data that we use for the final model evaluations. A variety of classification models from the scikit-learn [14] library were trained, including k-Nearest Neighbors, Decision Tree, Support Vector and Gaussian Naive Bayes. We also included ensemble models such as the Bagging Classifier, Random Forest, Extra Trees and Gradient Boosting. Ensemble methods are a machine learning technique that combines several base models in order to produce one optimal predictive model [15]. The basic idea is to combine the predictive power of many "weak learners" and in doing so the error can be dramatically reduced. For each model we use $k$-fold cross-validation to estimate the performance, ensuring the test data remains untouched. The idea behind $k$-fold cross-validation is to use training data to generate multiple mini train-test splits and use these splits to tune the model. In standard $k$-fold cross-validation, the data is partitioned into $k$ subsets, called folds. The algorithm is iteratively trained on $(k-1)$ folds while using the remaining fold as the test set. Figures 6 and 7 displays results of training models for cavity identification and fault classification, respectively. Note that the ensemble models generally outperform the others.

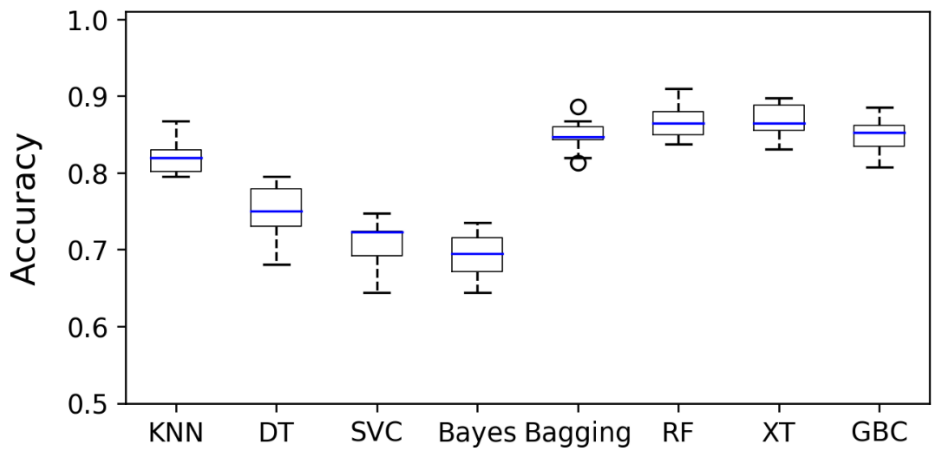

FIG. 6: Boxplots showing accuracy scores from a 10-fold cross-validation analysis of several algorithms for cavity identification. Ensemble models (four rightmost boxplots) exhibit the best performance.

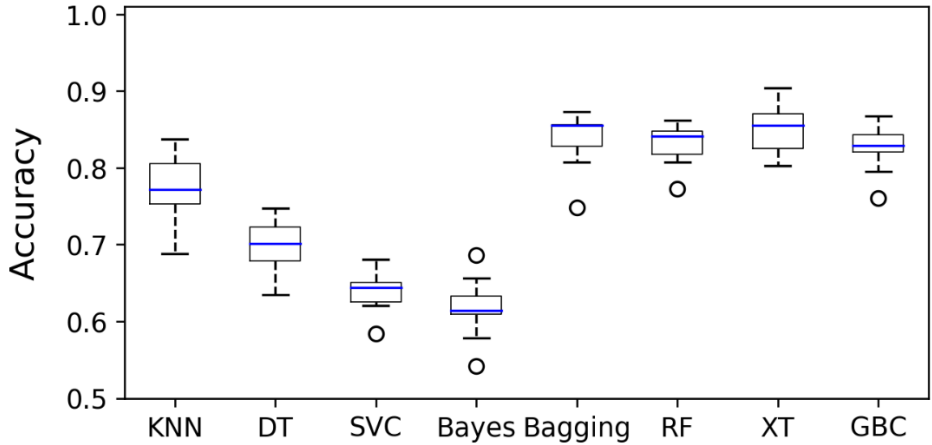

FIG. 7: Boxplots showing accuracy scores from a 10-fold cross-validation analysis of several algorithms for cavity fault classification. Ensemble models (four rightmost boxplots) exhibit the best performance.

### D. Model selection and hyperparameter tuning

The Random Forest Classifier was chosen to model both the cavity identification and fault classification because of its good performance (see Figs. 6 and 7) but also for its robustness against overfitting – a concern when dealing with smaller data sets. The next step is to optimize the model through hyperparameter tuning. Hyperparameters are parameters that are not directly learned but rather are passed as arguments to the constructor of the estimator classes. Rather than do a manual search of the parameter space, we invoke GridSearchCV to automate the process of scanning over specified parameter ranges. Specifically, in the case of the Random Forest we scan (*n_estimators, criterion,* and *bootstrap*). Once again, we avoid evaluating the model on the test data and take advantage of 10-fold cross-validation to estimate model performance, thereby keeping the test set as a truly unseen dataset for final model evaluation. The hyperparameters corresponding to the best accuracy score define the final model and are given in Table 1.

TABLE 1: Results of hyperparameter tuning for each Random Forest model.

|  | **Cavity Identification** | **Fault Classification** |
|---|---|---|
| *n_estimators* | 300 | 800 |
| *criterion* | *gini* | *gini* |
| *bootstrap* | *False* | *False* |

### E. Performance metrics

There are a multitude of metrics that can be used to assess the performance of classification models. Initially the 10-fold cross-validation score (based on training data) is compared with predictions from the withheld test set. In particular, we want to avoid to overfitting, marked by a cross-validation score being much greater than the accuracy derived from test data. Table 2 summarizes the results and indicates that overfitting, and the inability of a model to generalize learning to data outside the training set, is avoided.

TABLE 2: Cross-validation and accuracy scores for each Random Forest model. The accuracy scores were generated by applying the model on the withheld test data set.

|  | Cavity Identification | Fault Classification |
|---|---|---|
| 10-fold cross-validation (%) | 87.97 ± 1.81 | 85.52 ± 3.65 |
| accuracy (test data) (%) | 87.94 | 87.66 |

### F. Accessibility

Making data publicly available is necessary to drive innovation and progress. In an effort to make data Findable, Accessible, Interoperable and Reusable (FAIR) [16], the feature matrix (2,375 instances × 192 features/instance) and associated cavity identification and fault classification labels used in this publication are publicly available [17]. Furthermore, to make the data more accessible to non-domain scientists we follow the guidance put forward by Ref. [18] and provide an associated datasheet as documentation. Likewise, the saved Random Forest models are also available [19]. Analogous to the concept of datasheets for datasets, we follow the recommendations of Ref. [20] and provide a "model card" for each model, which describes characteristics such as the type of model, intended use cases, and performance metrics.

### G. System implementation

Following a C100 cavity fault event, software automatically collects RF signals, saves the data to disk and passes it to the machine learning models. The entire workflow is summarized in Algorithm 1.

| Algorithm 1: Machine learning for online cavity and fault recognition |
|---|
| - Initialization<br>    o Import relevant libraries in python<br>    o Load trained classifier models for cavity identification and fault classification<br>    o Read-in the RF signals for current instance<br><br>For each cryomodule:<br>- Preprocessing<br>    o Check for missing waveforms and missing cavities (data not recorded)<br>    o Select the 4 representative waveforms (GMES, GASK, CRFP, DETA2) from each cavity, and aggregate<br>    o Standardize each waveform using 'z-score' method in Eq. (1)<br>- Feature Extraction<br>    o Fit an $AR(5)$ (Eq. (3)) model to each of the 32 waveforms<br>    o Extract the parameters $\Phi = \{\varphi^0, \varphi^1, \ldots, \varphi^5\}$ from each AR model and aggregate to form a $6 \times 32 = 192$ feature vector<br>- Inference<br>    o Apply trained cavity identification classifier to feature vector and obtain output<br>    o Apply trained fault classification classifier to feature vector and obtain output<br>    o Aggregate output for each cryomodule<br>End<br><br>- Report<br>    o Save and display machine learning results to the operators |

The outputs are stored in a database and presented to operators via a web-based RF waveform viewer. Results can be presented in a table, downloaded as a comma separated file or plotted graphically (see Fig. 8). Offline, a SME labels the event and these labels are used to compare the model performance and to add to the training data for additional re-training in the future.

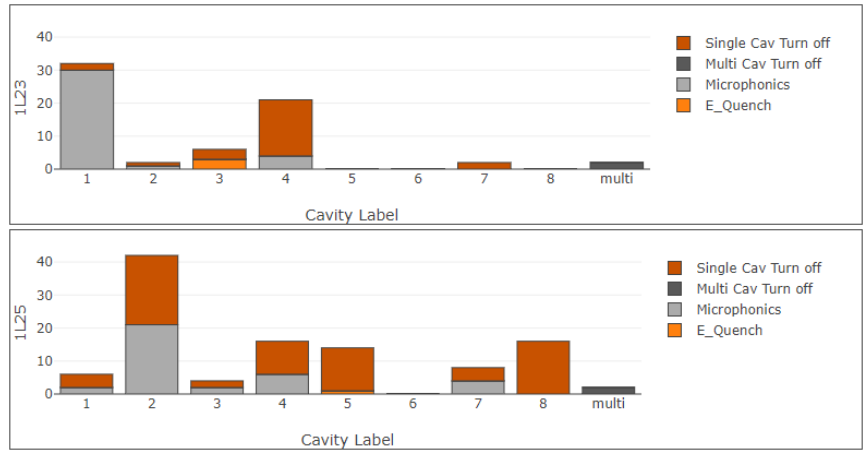

FIG. 8: Partial screenshot showing a graphical summary of machine learning model predictions. Each plot window provides information about an individual cryomodule. Fault types are color-coded and displayed for each of the eight cavities.

## VI. RESULTS

Machine learning models were deployed online and used to analyze C100 cavity fault events from March 10 to March 24, 2020 – until CEBAF operations ended prematurely due to COVID-19. During that two-week period 312 fault events were labeled by the machine learning models as well as a SME. The models identified the first cavity to fault and the type of fault with accuracies of 84.9% and 78.2%, respectively. The performance of the cavity identification model is consistent with the performance metrics in Table 2, however, the fault classification model underperforms. The models correctly predicted both the cavity and fault for 73.1% of the cases.

While these accuracies provide a general sense of how well the models are performing, a more insightful metric for classifier performance is the confusion matrix. A confusion matrix summarizes the percentage of correct and incorrect predictions by class. It provides an intuitive visualization for which classes are mislabeled (where the model is "confused"), thereby providing insight not only into the errors being made but, more importantly, the types of errors being made. A perfect model would show 0% on all off-diagonal elements and 100% along the diagonal. Figures 9 and 10 show the confusion matrix for the cavity and fault model, respectively.

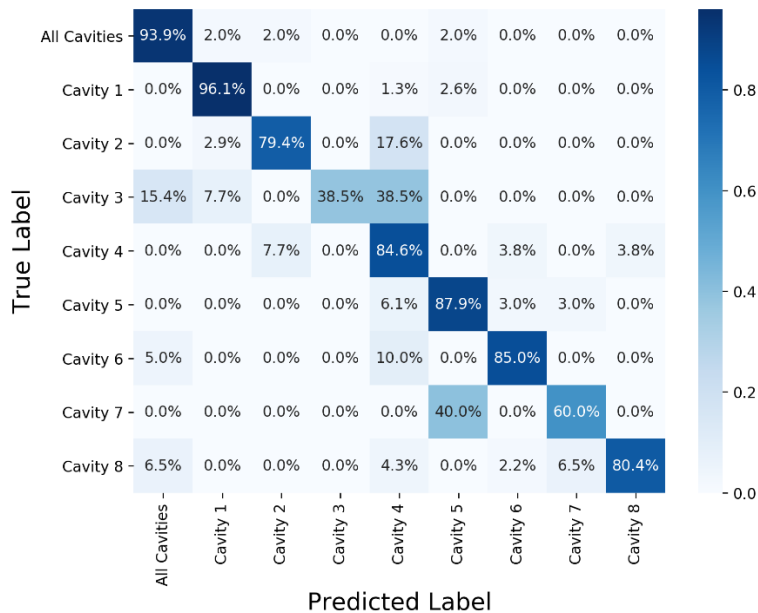

FIG 9: Confusion matrix showing performance of the deployed cavity identification model on 312 fault events compared to the labels provided by a subject matter expert.

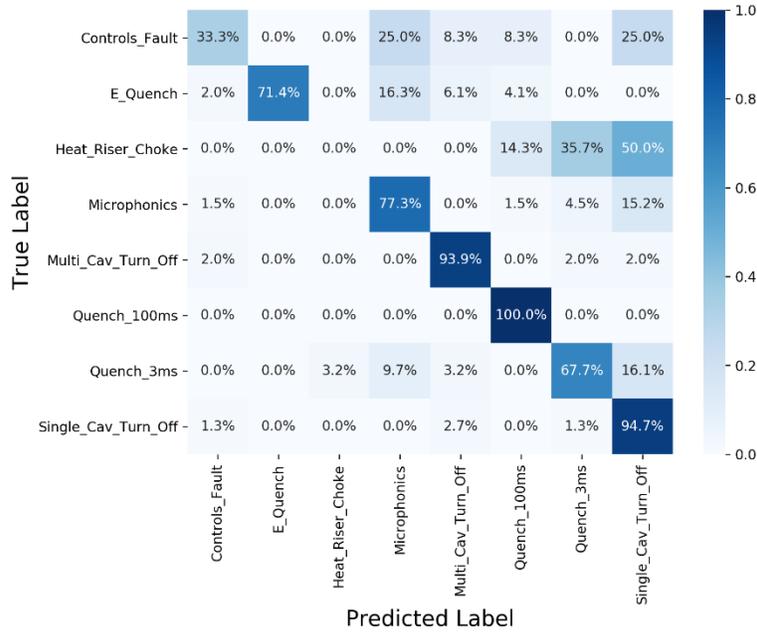

FIG. 10: Confusion matrix showing performance of the deployed fault classification model on 312 fault events compared to the labels provided by a subject matter expert.

Several observations are immediately apparent:
- The cavity model has particular difficulty identifying cavities 3 and 7. This is not surprising as these two cavities are the least represented in the training data (see Fig. 3).
- The fault model was unable to correctly classify heat riser choke faults. Similar to the previous point, this fault has the smallest number, by far, of training examples (see Fig. 4). However, there are additional reasons for the model's confusion with regard to this particular fault that will be discussed below.
- The fault model is easily confused by controls faults. Discussions with a system expert confirm that, with a few exceptions, it is typically difficult to understand the underlying mechanism of controls faults from the waveform data. This confirms that when the labeling is difficult, not surprisingly, it is often reflected in poor model performance.

During the 2-week model testing in March, there were several clusters of events where the models performed poorly and that warranted closer inspection. These are discussed briefly below.

- During the evening of March 16 eight heat riser choke faults were identified in cavity 2 in cryomodule 1L23, which the fault model failed to correctly label. These are relatively rare fault types and having a succession of them over a period of just a few hours represents anomalous behavior. On closer examination of the raw waveforms, the multiple faults is a product of operators not allowing the RF system sufficient time to settle from the previous fault before trying to recover the cavity. Further, the cavities that exhibited the characteristics of a heat riser choke (cavities 2 or 3) were not the cause of excess heat. Rather it was a thermal quench of cavity 1 which occurred several minutes earlier. The poor model performance may be attributed to the fact that the characteristics of the heat riser fault while recovering the cavity (as opposed to the event occurring during normal beam operation) were not well represented by the already limited training data for this fault type.
- On March 18 at 8:02 AM all of the C100 cryomodules tripped off simultaneously. This highly unusual event was quickly correlated to a central helium liquefier (CHL) compressor configuration change. In this case the machine learning cavity model failed to identify the correct cavity for each cryomodule trip. However, this is to be expected as the CHL event created conditions in the machine that were not represented in the training data and therefore the model struggled.
- It is worth noting that if these events are neglected, the cavity model accuracy increases to 86.6%.

Of the 312 events, there were 84 instances where the model disagreed with the labels (either with the cavity identification and/or the fault classification). These events were combined with 84 random events where the model agreed with both labels and submitted to a SME for relabeling. The accuracies of the models assume the labels are ground truth, yet we know there are errors associated with labeling. This was a blind test in that the SME was not given information about the previous labels. The results of the exercise were illuminating. Of the 84 events where the model predictions matched the initial labels, after re-labeling by the same SME, of the 168 labels (84 events × 2 labels/event) only 2 were labeled differently. On the other hand, of the 84 events where the model predictions differed from the initial labels, the SME re-labeled 58 of the 168 labels differently. This result underscores the fact that labeling for this task is difficult. It requires significant experience and intuition about SRF cavities operating with beam to understand the complex physical mechanisms that give rise to the data collected. And even then, faults can manifest themselves in subtle, but different, ways depending on details of the operating conditions which can vary from run to run and even day to day (i.e. vacuum conditions, gradient distribution, beam loss patterns, field emission).

When the new labels are compared to the machine learning model predictions, we find that the agreement significantly increases. Table 3 summarizes the accuracies of each model with each set of labels, original and re-labeled. Because only a subset (168 out of 312) of the events were relabeled, we make the assumption that events not relabeled (all of which agree with the models) would remain unchanged. This is reasonable given the 98.8% agreement of the SME when relabeling events where the model agreed the first time.

TABLE 3: Model accuracies as computed for the original set of labels and after a subset of events – including all that disagreed with the models – had been relabeled.

|  | Original | Re-Labeled |
|---|---|---|
| Cavity Identification | 84.9% | 88.1% |
| Fault Classification | 78.2% | 82.4% |
| Both | 73.1% | 78.2% |

As a result of the relabeling, the models more closely approach their anticipated performance (Table 2). However, more fundamentally, we find that when the models struggle it is because the subject matter expert struggles. This is not surprising. If the SME is less certain about particular kinds of events, it will manifest itself in the training data and therefore, what the model learns. The model is only as good as the training data it is exposed to.

While the aim is to develop machine learning models with accuracies as high as possible, it should be emphasized that for this to be a valuable tool for machine operators, the current level of performance is perfectly acceptable. The system provides near real-time information about C100 cavity faults that was previously unavailable. Operators and system experts alike, will utilize the information presented to look for trends over time that would necessitate a change in a cavity or cavities. In other words, the effectiveness of the system does not depend on the ability to accurately classify a single, standalone event.

When the models report consistent trips from the same cavity in a cryomodule and of the same fault type, this suggests that intervention by an operator or system expert is appropriate. Machine learning model predictions from the two-week physics run clearly point to several cavities that required attention (for this initial test, machine learning predictions were not presented to operators). The cavity identities, fault types and suggested interventions are summarized in Table 4. Together, these four cavities accounted for 19.6% of the faults during the physics run. Applying the appropriate intervention would have had a measurable effect on beam availability to users.

TABLE 4: Problematic cavities as identified by the machine learning models during a two-week test run in March 2020. The cavity identities, number of times the cavity tripped, fault-types and suggested interventions are given.

| | **Problematic Cavities** | | | |
|---|---|---|---|---|
| Cryomodule | 1L23 | 1L24 | 1L26 | 2L26 |
| Cavity Identification | 1 | 6 | 8 | 4 |
| Number of Trips | 18 | 15 | 23 | 5 |
| Fault Classification | Quench_3ms | Quench_100ms | E_Quench | Microphonics |
| Intervention | lower gradient | lower gradient | lower gradient | tune cavity |

## VII. FUTURE WORK

Initial experience with a prototype machine learning system has provided insights into opportunities to further advance this work. These include:

- Use unsupervised machine learning techniques to identify fault types. At present, a SME determines the fault classes. Therefore, when presented with unfamiliar pattern in the data, is raises the question of whether it is a variant of a known fault type or if it represents a completely new fault type. Invoking clustering algorithms [21] may provide a data-driven method for identifying the number of fault types represented in the data, and potentially reveal new fault types.
- Replace current machine learning models with their deep learning counterparts. Deep learning is a sub-field of machine learning which is based on learning successive layers of increasingly meaningful representations of the data. The primary advantage of methods based on learning data representations is that it avoids the computationally costly feature extraction step, thereby bypassing a bottleneck in our current workflow. Rather than explicitly compute features, the recorded data serves as the input and the model learns the necessary features for inference.
- Initial studies suggest machine learning can extract information in the signals preceding the fault for prediction, such that it may be possible to design future SRF cavities and their associated control systems with the ability to react quickly enough to avoid the fault altogether [22]. A natural extension of this work is to develop a data pipeline, workflow, and models to stream continuous C100 RF data and provide real-time predictions for impending faults. This represents a critical step towards developing systems that could anticipate and then apply corrective actions to avoid the fault.

## VIII. SUMMARY

We have described an implementation of machine learning models at CEBAF to automate the task of classifying C100 SRF cavity faults. The success of these models is a direct consequence of collecting the appropriate data. The importance of having the right data, at the right time cannot be overemphasized. This would not have been possible without the capabilities of a digital LLRF system, together with a specially developed data acquisition system. Significant investment was necessary to get a functional DAQ, requiring the coordinated effort of multiple groups at the lab. In addition to collecting information-rich data, an equally important contribution was the process of analyzing and manually labeling the data. Although time consuming, the labeling task is an effective way to identify trends and to strategically deploy mitigations to problematic cryomodules and to implement design changes in interlocks. By leveraging machine learning to automate this process, we now have for the first time, that information available in near real-time.

The next physics run will provide an opportunity to translate model predictions to actionable information for machine operators. For instance, establish a list of guidelines that connects model predictions with operator actions (e.g. "if fault type $X$ happens $Y$ times in the next $Z$ minutes, reduce cavity gradient by $N$ MV/m"). This represents the natural next step for this work, however, initial efforts were hampered by the premature shutdown of CEBAF. In principal, this could be automated in software, however it is important in this nascent stage of development for operators to gain trust in the machine learning system and allow a human operator to take the decisive action.

## ACKNOWLEDGEMENTS

This work is supported by Jefferson Science Associates, LLC under U.S. DOE Contract No. DE-AC05-06OR23177. The authors gratefully acknowledge Dr. Andrei Seryi, Dr. Todd Satogata and Dr. Yves Roblin for their encouragement and support of this work.

# APPENDIX A: RF SIGNALS AND DESCRIPTION

Table A1 gives a list of the 17 RF signals recorded by the waveform harvester during each C100 cavity trip, along with a brief description. The 4 signals used for feature extraction and subsequent training of the machine learning models are: (GMES, GASK, CRFP, DETA2).

TABLE A1: List of the 17 RF signals recorded during a cavity trip along with a brief description of each.

| Signal Name | Description |
| --- | --- |
| IMES | probe in phase voltage magnitude (analog-to-digital converter (ADC) counts) |
| QMES | probe quadrature voltage magnitude (ADC counts) |
| GMES | measured gradient (MV/m) |
| PMES | measured phase (degrees) |
| IASK | drive signal in phase voltage magnitude (ADC counts) |
| QASK | drive voltage quadrature voltage (ADC counts) |
| GASK | $\sqrt{IASK^2 + QASK^2} \times (10 / \text{ADC maximum})$ |
| PASK | $\tan^{-1}(IASK/QASK)$ |
| CRFP | forward power (kW) |
| CRFPP | forward power measured phase (degrees) |
| CRRP | reflected power (kW) |
| CRRPP | reflected power phase (degrees) |
| GLDE | gradient error (MV/m) |
| PLDE | phase error (degrees) |
| DETA2 | detune angle (CRFPP − PMES − TDOFF) |
| CFQE2 | cavity frequency error (degrees) |
| DFQES | discriminator frequency error |